\documentclass{aa}
\usepackage{graphicx,txfonts}
\usepackage{natbib}
\bibpunct{(}{)}{;}{a}{}{,}

\def\cm-2{cm$^{-2}$}

\def\msun{M$_{\odot}$}

\def\ein{{\it Einstein}}
\def\chandra{{\it Chandra}}
\def\swift{{\it Swift}}

\def\xmm{{\it XMM-Newton}}
\def\m31{M~31}
\def\me33{M~33}
\def\n12b{\hbox{M31N~2007-12b}}

\newcommand{\ergs}[1]{$\times 10^{#1}$ \hbox{erg s$^{-1}$}}
\newcommand{\oergs}[1]{$10^{#1}$ erg s$^{-1}$}
\newcommand{\hcm}[1]{$\times 10^{#1}$ cm$^{-2}$}
\newcommand{\ohcm}[1]{$10^{#1}$ cm$^{-2}$}

\newcommand{\nh}{\hbox{$N_{\rm H}$}}

\newcommand{\cts}{ct s$^{-1}$}

\begin{document}

   \title{Nova M31N 2007-12b: Supersoft X-rays reveal an intermediate polar?\thanks{
   Based on observations obtained with {\it XMM-Newton}, an ESA Science Mission 
    with instruments and contributions directly funded by ESA Member
    States and NASA}
}

   \subtitle{}

   \author{W.~Pietsch\inst{1} \and 
           M.~Henze\inst{1} \and           
           F.~Haberl\inst{1} \and
	   M.~Hernanz\inst{2} \and
	   G.~Sala\inst{3} \and
	   D.H.~Hartmann\inst{4} \and
	   M.~Della Valle\inst{5,6,7} 
          }
\institute{Max-Planck-Institut f\"ur extraterrestrische Physik, Giessenbachstra\ss e, 
           D-85741 Garching, Germany \\
	   \email{wnp@mpe.mpg.de}
	   \and
	   Institut de Ci\`encies de l'Espai (CSIC-IEEC), Campus UAB, Fac. Ci\`encies,
	   E-08193 Bellatera, Spain 
	   \and
	   Departament de F\'isica i Enginyeria Nuclear, EUETIB  
           (UPC-IEEC), Comte d`Urgell 187, E-08036 Barcelona, Spain
	   \and
	   Department of Physics and Astronomy, Clemson University, Clemson, SC
	   29634-0978, USA
	\and European Southern Observatory (ESO), D-85748 Garching, Germany
	\and INAF-Napoli, Osservatorio Astronomico di Capodimonte, Salita Moiariello 16, I-80131 Napoli, Italy
	\and International Centre for Relativistic Astrophysics, Piazzale della Repubblica 2, I-65122 Pescara, Italy
           }
     

   \date{Received / Accepted }

	\abstract
{In the central part of \m31, a high number of optical novae can be targeted within the
field of view of the \xmm\ EPIC and \chandra\ HRC-I X-ray detectors. A special monitoring
program of the area has allowed us to investigate supersoft emission of individual novae in
detail and perform a statistical analysis of the sample.}  
{For the He/N nova \n12b, we aimed to constrain
the time of appearance of a supersoft source (SSS) and the duration of the SSS state as
well as determine the spectral and time variability while the source was bright.} 
{We analyzed \xmm\ EPIC and \chandra\ HRC-I
observations of our monitoring program performed at intervals of ten days and added results of a
\xmm\ target of opportunity observation and \swift\ XRT  observations. We performed
source detection, determined long-term time and spectral variation of \n12b, and searched
for shorter-term time variability in the individual observations when the source was
bright, using fast Fourier and folding techniques to analyze periodicities.} 
{The SSS
emission started between 21 and 30~d after the optical outburst and ended between 60 and
120~d after outburst, making \n12b\ one of the few novae with the shortest SSS phase known.
The X-ray spectrum was supersoft and can be fitted with a white dwarf (WD) atmosphere model
with solar abundances absorbed by the Galactic foreground. The temperature of the WD
atmosphere seems to increase at the beginning of the SSS phase from $\sim$70 to $\sim$80~eV. 
The luminosity of \n12b\ during maximum was at the Eddington limit of a massive WD and 
dropped by $\sim$30\% in the observation 60~d after outburst. The radius of the emission
region is $\sim6\times 10^{8}$~cm. In the four bright state 
observations, we detected a stable 1110 s pulsation, which we interpret as the WD
rotation period. In addition, we detect dips in three observations that might represent a 
4.9~h or 9.8~h binary period of the system.} 
{Nova envelope models with $\la$50\% mixing between solar-like accreted material and the 
degenerate core of the WD can be used to describe the data.
We derive a WD mass of 1.2~M$_{\sun}$, as well as an ejected 
and burned mass of $2.0\times 10^{-6}~M_{\sun}$ and $0.2\times 10^{-6}~M_{\sun}$, respectively. 
The observed periodicities indicate that nova \n12b\
erupted in an intermediate polar (IP) system. The WD photospheric radius seems to be 
larger than expected for a non-magnetic WD but in the range for 
magnetic WDs in an IP system. }	
\keywords{Galaxies: individual: \m31 -- 
          novae, cataclysmic variables -- 
	  stars: individual: Nova M31N 2007-12b --
          X-rays: galaxies -- X-rays: binaries 
}

\maketitle

\section{Introduction}
Outbursts of classical novae (CNe) are caused by a thermonuclear runaway on the surface of
a white dwarf (WD) in a cataclysmic variable (CV), i.e. a close binary system with
transfer of material from the companion star to the WD \citep[see
e.g.][]{1998ApJ...494..680J,2005ApJ...623..398Y,2008clno...77.....S}. Ignition takes place in the accreted
envelope and as a consequence, the envelope expands and causes the brightness of the star
to increase. While a fraction of the envelope is ejected, part of it remains in steady
nuclear burning on the WD surface. This area radiates as a supersoft X-ray source (SSS),
which can be identified with the spectrum of a hot ($T_{eff}: 10^{5}-10^{6}$~K) WD
atmosphere as soon as the soft X-rays can penetrate the ejected envelope
\citep[][]{1991ApJ...373L..51M}. The duration of the SSS phase is shorter for more massive
WDs. It also depends on the metallicity of the envelope. The time of appearance of the SSS
is determined by the mass ejected in the outburst and the ejection velocity \citep[for
model calculations see e.g.][]{2006ApJS..167...59H}. 
 
Models \citep[see e.g.][]{2006ApJS..167...59H,2010ApJ...709..680H} have been 
developed to explain these smooth SSS light curves. However, specifically at the start of the SSS 
phase several Galactic novae have shown strong -- possibly periodic -- time variability 
\citep[see e.g.][]{2007ApJ...659L.153H,2009AJ....137.4160N,2009ATel.2025....1B}. 
Pulsations of the SSS emission may indicate the WD rotation 
period \citep[see e.g.][]{2010MNRAS.405.2668D}. 
However, until now it has mainly been inferred from optical data that several novae 
exploded in intermediate polar (IP) systems \citep{2002AIPC..637....3W}.
IPs (often also called DQ Herculis stars after the prototype of the class) 
are cataclysmic variables containing an 
accreting, magnetic, rapidly rotating WD. In contrast to the polar (AM~Herculis) 
systems, the orbital and rotation period of the WD are not synchronized. The regular 
pulsations in IPs seen in optical and X-rays are explained by magnetically 
channeled accretion onto the poles of the WD within
a truncated disk \citep[for a review, see][]{1994PASP..106..209P}.

Recurrent novae (RNe) are CNe that have more than one recorded outburst. RNe
are specifically interesting as they have been discussed as one of the most likely
progenitors of type Ia supernovae \citep[see e.g. reviews by][]{2008ASPC..401...31A,2008ASPC..401..150K}.  

The optical outburst of nova \n12b\ was detected independently by several groups 
\citep[see entries on the Central Bureau for Astronomical Telegrams CBAT M31
(Apparent) Novae Page\footnote{http://www.cfa.harvard.edu/iau/CBAT\_M31.html} and][]
{2007ATel.1324....1L}. The time of outburst was constrained to better than a day by 
Nishiyama 
and Kabashima using unfiltered images (first detection with 16.1 mag and last
non-detection with lower limit of 18.9 mag on 2007 December 9.528 and 8.574 UT,
respectively). Throughout the paper, we use 2007 December 9.0 UT as the time of the outburst. 
\citet[][]{2010arXiv1003.1720K} present a Gunn-g band light curve. The nova was first detected 
2.7~days after outburst and monitored during the following 5~days. They give a rate of decline from maximum by one 
and two magnitudes of 3.5 and $>$5~days, respectively.    
\citet[][hereafter BDS2009]{2009ApJ...705.1056B} report broadband i', V, B, and narrow-band H$\alpha$ photometry 
that began 5.9~days post-outburst and then continued for 23~days producing a fast decay. 

Nishiyama and Kabashima determined the position of the nova to 
RA = 00$^{\rm h}$43$^{\rm m}$19\fs94$\pm$0\fs01,
Dec = $+$41\degr13\arcmin46\farcs6$\pm$0\farcs1 (J2000). Based on observations with the 
2-m Liverpool
Telescope, BDS2009 confirmed the position (end figures 19\fs97$\pm$0\fs01 
in RA and 46\farcs3$\pm$0\farcs1 in Dec) applying an astrometric solution to a 2-m Liverpool Telescope 
Sloan i'-band image using stars from the Two Micron All Sky Survey (2MASS) Catalogue 
\citep[][]{2003tmc..book.....C}.
They showed that the position of \n12b\ is inconsistent with
the nearby nova \hbox{M31N 1969-08a}. BDS2009 also analyzed {\it Hubble Space Telescope} 
observations of the pre-outburst
location of \n12b. They found a stellar source positionally coincident with \n12b\ with magnitude and color very
similar to the Galactic recurrent nova RS Ophiuchi in quiescence, where the red giant dominates the
emission. BDS2009 propose this source as the first nova progenitor system identified in \m31.

From optical spectroscopic data obtained on 2007 December 15, 
the nova is classified as He/N following the classification scheme of 
\citet[][]{1992AJ....104..725W},
with a full width at half maximum (FWHM) of the H$\alpha$ line of $\sim$4500 km s$^{-1}$ 
\citep[][BDS2009]{2007ATel.1332....1S}. \citet[][]{2008IAUC.8907....3R} also find
notably broad lines for a nova in a spectrum taken 2007 December 12. They measure a full width at 
zero maximum (FWZM) of the H$\alpha$ line of $\sim$6800~km~s$^{-1}$. 

\citet[][]{2008ATel.1360....1K} reported the discovery of a new X-ray source 
with the \swift\ satellite \citep{2004ApJ...611.1005G} X-ray telescope (XRT) 
on 2008 Jan 13 (36~days after outburst) at 
\hbox{RA = 00$^{\rm h}$43$^{\rm m}$20\fs2,}
\hbox{Dec = $+$41\degr13\arcmin48\arcsec}~(J2000) with a 90\% error
radius of 4.5 arcsec consistent with the position of the optical nova \n12b. No
source was present at the position in {\it Swift} observations on 2007 Dec 16 (day~8) and Dec
30 (day~21). The emission of the new source was supersoft (all 48 X-ray photons with 
energies below 0.8~keV). The source is interpreted as a supersoft transient
associated with \n12b\ detected 36~d after the optical outburst. 

BDS2009  re-analyzed {\it Swift} observations of the field. 
No source was detected in
a {\it Swift} observation on 2008 May 13 (day~170, 3$\sigma$~upper limit about a factor of eight
below the outburst count rate). They estimate the mass of the WD in the system to be 
most likely greater than
1.3 \msun\ and quote that the optical light curve, spectrum, and X-ray behavior are
consistent with that of a RN. According to BDS2009, \n12b\ shows the greatest similarities to the supposed 
Galactic RN V2491~Cygni \citep[of the U Sco variety,][]{2010MNRAS.401..121P}. This classification 
would indicate that it is a short period system (orbital period of hours to shorter 
than a few days) and contradict the optical 
identification in quiescence of a red giant system \citep[see
above, orbital period $>$100~days to a year,][]{2008ASPC..401...31A}.  

Starting in 2006, we carried out a dedicated optical and X-ray monitoring program of novae 
and SSSs in the central area of 
\m31\footnote{See http://www.mpe.mpg.de/$\sim$m31novae/xray/index.php}, which also covered \n12b. 
These observations were supplemented with a \xmm\ target of opportunity (ToO) observation of the \n12b\ field 
in July 2008 targeted by ourselves to enable the length of the SSS phase to be more accurately constrained. 
First results from these observations
-- including the detection of a 1100~s periodicity -- were reported by 
\citet{2010AN....331..187P} and \citet{2010ApJ...717..739O}.
In this paper, we discuss the observations in detail concentrating on the X-ray
spectra and both the short- and
long-term variability of the supersoft X-ray counterpart of \n12b.

\section{Observations and data analysis}
\subsection{Optical data}\label{obs_opt}
\begin{figure*}
  \resizebox{\hsize}{!}{\includegraphics[angle=90,clip]{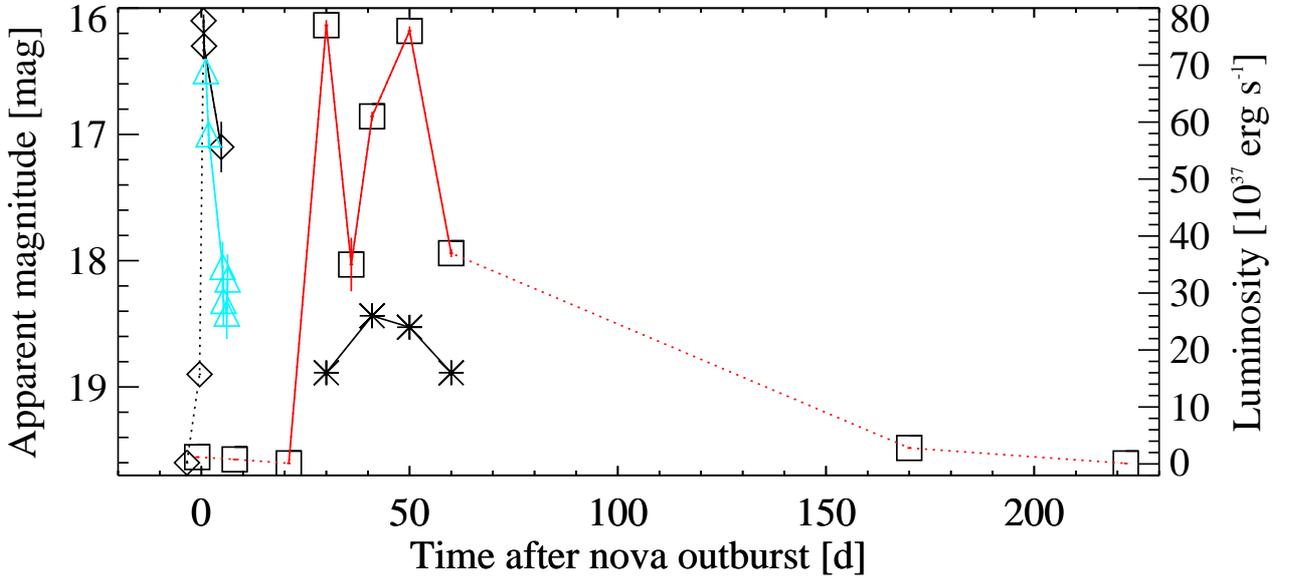}}
     \caption[]{
     Optical and X-ray light curve of M31N 2007-12b assuming a time of
     nova outburst of 2007 December 9.0 UT. 
     Optical white light
     (diamonds) and R filter (triangles) apparent magnitudes are plotted using data from the CBAT M31
     (Apparent) Novae Page and additional Super-LOTIS data (see Sect.~\ref{obs_opt}, left y-axis).
     X-ray luminosities and upper limits derived for an absorbed black-body spectrum, are 
     plotted as squares (see 
     Table~\ref{observations}, right y-axis).
     Detections are connected with solid lines, and upper limits with dotted lines. 
     Statistical errors in the \xmm\ EPIC pn data points are smaller than the symbol size.
     For the four \xmm\ bright state observations, we also give the bolometric 
     luminosities derived for the WD atmosphere model with solar abundances
     (stars, see Table~\ref{spectra}).  
     }
    \label{nova_lc_opt_xray} 
\end{figure*}

In addition to the optical observations reported in the introduction, we obtained optical photometric 
data from Super-LOTIS \citep[Livermore Optical Transient Imaging System,][]{2008AIPC.1000..535W}, a
robotic 60 cm telescope equipped with an E2V CCD (2k$\times$2k, pixel scale 0.496\arcsec/pixel) 
located at Steward Observatory, Kitt
Peak, Arizona, USA. Our data reach a
limiting magnitude of $\sim$19 mag in Johnson R. The data were reduced in a semi-automatic routine and 
photometrically calibrated with the help of the observations of \m31\ in the Local Group Galaxy
Survey \citep[LGGS,][]{2006AJ....131.2478M}. We achieve typical accuracies of 0.25 mag for
Super-LOTIS photometry averaged over the whole magnitude range.
From sets of images at 2007 December 14.1, 14.3, 15.1, and 15.3, we determine R magnitudes for \n12b\ of 
18.05, 18.32, 18.42, and 18.15, respectively.
These data are included in the optical light curve of \n12b\ (Fig. \ref{nova_lc_opt_xray}), which also shows
measurements reported on the CBAT M31
(Apparent) Novae Page and those of \citet[][]{2007ATel.1324....1L} and \citet[][]{2007ATel.1329....1P}.

\subsection{X-ray data}\label{obs_x-ray}
\begin{table*}
\caption{\label{observations}X-ray observations of the 
\n12b\ field around and after the optical nova outburst.}
\centering
\begin{tabular}{lllrrrrrr}
\hline\hline
\multicolumn{1}{c}{Observatory} &\multicolumn{1}{c}{Instr.} &
\multicolumn{1}{c}{ObsID} & \multicolumn{1}{c}{Obs. start} &
\multicolumn{1}{c}{Exposure} & \multicolumn{1}{c}{$\Delta t$\tablefootmark{a}} & 
\multicolumn{1}{c}{Count rate\tablefootmark{b}} & 
\multicolumn{1}{c}{L$_{\rm X}$\tablefootmark{c,b}} & \multicolumn{1}{c}{Comment} \\ 
& & & (UT) & (ks) & (d) & (ct ks$^{-1}$) & (\oergs{37})\\
\hline
\chandra &   HRC-I& 8529       & 2007-12-07.57 & 19.1 &  -1 & $     <1.3 $ &   $     <1.2$ & off\\
\swift   &     XRT& 00031027004& 2008-12-16.78 &  3.9 &   8 & $     <3.9 $ &   $     <9.2$ & off\\
\chandra &   HRC-I& 8530       & 2007-12-17.49 & 20.2 &   8 & $     <0.9 $ &   $     <0.8$ & off\\
\xmm     & EPIC pn& 0505720201 & 2007-12-29.57 & 22.3 &  21 & $ 1.0\pm0.4$ &   $     0.11$ & faint\\
\swift    &     XRT& 00031027005& 2008-12-30.02 &  4.0 &  21 & $     <3.4 $ &   $     <8.0$ & off or faint\\
\xmm     & EPIC pn& 0505720301 & 2008-01-08.29 & 22.1 &  30 & $   260\pm3$ &   $       77$ & bright, dip\\
\swift    &     XRT& 00031027006& 2008-01-13.74 &  4.0 &  36 & $    15\pm2$ &   $       35$ & bright\\
\xmm     & EPIC pn& 0505720401 & 2008-01-18.63 & 18.2 &  41 & $   395\pm5$ &   $       61$ & bright, dip\\
\xmm     & EPIC pn& 0505720501 & 2008-01-27.94 & 17.3 &  50 & $   479\pm5$ &   $       76$ & bright, dip\\
\xmm     & EPIC pn& 0505720601 & 2008-02-07.21 & 17.4 &  60 & $   291\pm4$ &   $       37$ & bright\\
\swift    &     XRT& 00037719001& 2008-05-26.71 &  4.9 & 170 & $     <2.3 $ &   $     <5.4$ & off\\
\xmm     & EPIC pn& 0560180101 & 2008-07-18.26 & 17.4 & 222 & $     <1.5 $ &   $     <0.1$ & off\\
\hline
\end{tabular}
\tablefoot{
\tablefoottext{a}{time after optical outburst (assumed 2007 December 9.0 UT)}
\tablefoottext{b}{upper limits are 3$\sigma$.}
\tablefoottext{c}{0.15--1.0~keV absorption corrected luminosity assuming an absorbed 
black-body spectrum as determined in the spectral fit (see Table \ref{spectra}).  
For observations where spectral fitting was not possible, we assumed a black-body
spectrum with 
\nh = 1.0\hcm{21} and a temperature $kT=60$~eV. We use a distance of \m31\ of 
780~kpc \citep{1998AJ....115.1916H,1998ApJ...503L.131S} throughout the paper.}
}
\end{table*}

A \chandra\ and \xmm\ \m31\ project monitored novae in the central area of \m31\ 
from November 2007 to February 2008 with observations
separated by ten days. It covered the outburst of \n12b. Details of observations used are summarized in Table
\ref{observations}. The table lists the telescope and instrument used,  the observation
identifications (ObsIDs), the observation start times, observation exposure times, and times
after optical outburst of the nova. We give count rates, X-ray luminosity in the 0.15--1.0~keV 
band, and comment on the X-ray brightness.
We also report the \swift\ observations of the field
and the additional \xmm\ ToO observation (ObsID 0560180101) pointed at the
position of \n12b. 

The \xmm\ \citep{2001A&A...365L...1J} observations were performed with the 
EPIC instruments \citep{2001A&A...365L..18S,2001A&A...365L..27T} in the full frame mode and
filters thin and medium for pn and MOS detectors, respectively.  For the \xmm\ data
analysis, we used XMMSAS v10.0 tasks and calibration files that had been made 
public before 2010 August, which
included the latest EPIC pn energy redistribution file (v11). 
For source detection, we rejected times with high
background. Sources were simultaneously searched for in 15 images binned to
2\arcsec$\times$2\arcsec\ in the energy bands \hbox{0.2--0.5,} \hbox{0.5--1,} \hbox{1--2,} \hbox{2--4.5,} and \hbox{4.5--12~keV,} 
using single and double pixel events for pn (only single pixel events for the 0.2--0.5~keV band)
and single to four pixel events for the two MOS detectors. The astrometric accuracy was
improved by a pointing offset correction using the catalogue of \chandra\ HRC-I sources by
\citet[][]{2002ApJ...578..114K}.  For a more detailed description, we refer to \citet[][]{2010A&A...523A..89H}. 
In this way, the remaining systematic positional error (which dominates for bright
sources) can be reduced to $\sim$0.5\arcsec\ (1$\sigma$).  For the timing and spectral analysis
of \n12b, we only used EPIC pn data since the source was on or close to CCD boundaries in
the MOS detectors and because of the higher sensitivity of the pn detector compared to the
MOS detectors in the soft energy band. For this analysis, we also accepted times of higher 
high-energy background as they are not affecting as much the count rates in the soft X-ray bands.  
To reduce the background we only
selected single-pixel events that dominate the energy range covered by SSSs.
In the timing analysis, we corrected the event arrival times to the solar system barycenter
and created background-corrected light curves. The count rate and upper limits in ObsIDs
0505720201 and 0560180101 were determined with emldetect set to position fixed mode using as
an input source list the sources detected in the observation and in addition the position of
\n12b.

M31N~2007-12b was not detected in the \chandra\ \citep{2002PASP..114....1W} HRC-I 
\citep{2000SPIE.4012...68M} observations of our monitoring. For these observations, we
determined upper limits to the count rate of a source at the nova position. 

We analyzed the observations using mission-dependent software as well as the HEAsoft
package v6.3, including the spectral analysis software XSPEC v12.3.1. For the XSPEC models,
we used the T\"ubingen-Boulder ISM absorption (\texttt{TBabs} in XSPEC) model together
with the photoelectric absorption cross-sections from \citet{1992ApJ...400..699B} and ISM
abundances from \citet{2000ApJ...542..914W}. 

\subsection{Ultraviolet data}\label{obs_uv}
\begin{table*}
\caption[]{\xmm\ OM and \swift\ UVOT observations of the 
\n12b\ field around and after the optical nova outburst.}
\centering
\begin{tabular}{lllrrrrrr}
\hline\hline
\multicolumn{1}{c}{Observatory} &\multicolumn{1}{c}{Instr.} &
\multicolumn{1}{c}{ObsID} & \multicolumn{1}{c}{Obs. start} &
\multicolumn{1}{c}{Filters} &
\multicolumn{1}{c}{Exposure} & \multicolumn{1}{c}{$\Delta t$\tablefootmark{a}} & 
\multicolumn{1}{c}{Brightness\tablefootmark{b}} \\ 
& & & (UT) & & (s) & (d) & (mag)\\
\hline
\swift   & UVOT & 00031027002& 2008-12-02.64 &    U & 1002 &  -6 & $ >19.8 $ \\
\swift   & UVOT & 00031027003& 2008-12-03.25 & UVW1 &  375 &  -6 & $ >19.3 $ \\
         &      &            &               & UVM2 &  399 &  -6 & $ >19.5 $ \\
         &      &            &               & UVW2 &  399 &  -6 & $ >19.6 $ \\
\swift   & UVOT & 00031027004& 2008-12-16.78 & UVM2 & 3853 &   8 & $  17.82\pm0.04 $ \\
\xmm     &   OM & 0505720201 & 2007-12-29.57 & UVW1 & 4800 &  21 & $  19.92\pm0.03 $ \\
\swift   & UVOT & 00031027005& 2008-12-30.02 &    U & 3977 &  21 & $  19.77\pm0.18 $ \\
\xmm     &   OM & 0505720301 & 2008-01-08.29 & UVW1 & 4720 &  30 & $  20.27\pm0.16 $ \\
\swift   & UVOT & 00031027006& 2008-01-13.74 & UVM2 &  841 &  36 & $ >19.9 $ \\
\xmm     &   OM & 0505720401 & 2008-01-18.63 & UVW1 & 3500 &  41 & $  20.62\pm0.05 $ \\
\xmm     &   OM & 0505720501 & 2008-01-27.94 & UVW1 & 3300 &  50 & $ >20.5 $ \\
\xmm     &   OM & 0505720601 & 2008-02-07.21 & UVW1 & 3680 &  60 & $ >20.5 $ \\
\hline
\end{tabular}
\label{uv_observations}
\tablefoot{
\tablefoottext{a}{time after optical outburst (assumed 2007 December 9.0 UT)}
\tablefoottext{b}{\swift\ magnitudes in UVOT photometric system \citep{2008MNRAS.383..627P}, 
for \xmm\ OM instrumental magnitudes}
}
\end{table*}
During the \xmm\ and \swift\ X-ray observations, \n12b\ was in the field of view of the 
\xmm\ optical monitor \citep[OM][]{2001A&A...365L..36M} and the \swift\ ultraviolet/optical 
telescope UVOT. Table~\ref{uv_observations} indicates the data observatory, instrument, ObsID, 
filter used, exposure time on \n12b, and its brightness or upper limit. No source is 
detected at the \n12b\ position in the U and ultraviolet (UV) filters six days before the 
nova outburst. The \swift\ observation taken eight days after the outburst already shows a bright 
source at the nova position in the combined UVM2 filter image (166--268~nm), which does not vary 
significantly in brightness between the four individual images of about equal exposure
distributed over $\sim$5~h. In the \swift\ UVM2 filter observation 24~d later, the source 
had faded by more than 2 mag and was no longer detectable. The OM UVW1 filter (240--360~nm)
observations only start 21 d after the outburst. \n12b is clearly detected. After 30~d, 
in this filter the source also faded below detectability. 

\section{Results}\label{results}
\begin{figure*}
   \resizebox{\hsize}{!}{\includegraphics[]{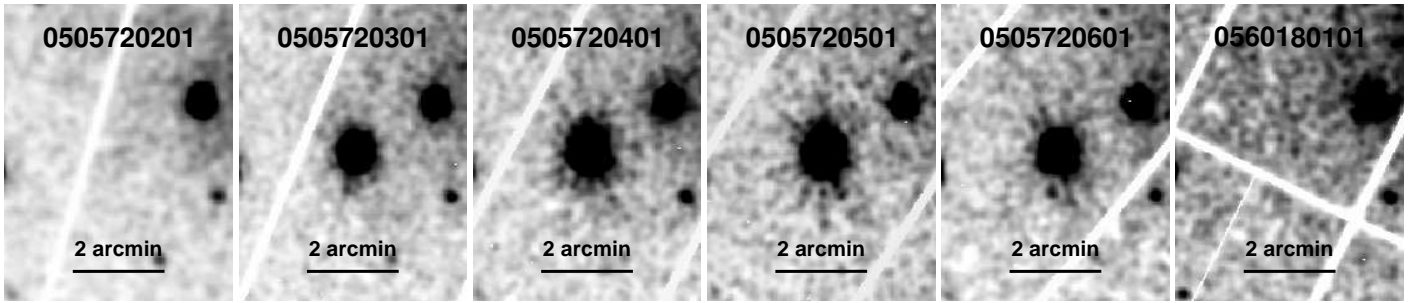}}
     \caption[]{
     \xmm\ EPIC pn images (0.2--1~keV) of the area around M31N 2007-12b -- from 
     left to right -- 21, 30, 41, 50, 60, and 222~days after the optical nova outburst,
     respectively (see Table \ref{observations}).
     }
    \label{nova_image_xray} 
\end{figure*}

In the optical, \n12b\ displayed a rapid rise to the outburst maximum (more than 2.5 mag in  
observations without any filter in one day, according to Nishiyama and Kabashima). 
Using \citet[][]{2007ATel.1324....1L} and Super-LOTIS data, the decay time by 2 mag 
in the Johnson R band (t$_{\rm 2R}$) can be constrained to  \hbox {5~d $<$  t$_{\rm 2R}$ $<$ 8~d.} 
Therefore we can classify   \n12b\   as a very fast nova \citep[following the speed class
scheme of][]{1964gano.book.....P}.
 
\begin{table}
\caption[]{Position of the SSS correlating with \n12b.} 
\centering
\begin{tabular}{lrrr}
\hline\hline
\multicolumn{1}{c}{\xmm}  &
\multicolumn{2}{c}{SSS position} & \multicolumn{1}{c}{Distance\tablefootmark{a}}\\ 
\multicolumn{1}{c}{ObsID} & RA (J2000) & Dec (J2000) & (")\\
\hline
0505720301& 00$^{\rm h}$43$^{\rm m}$19\fs94 & $+$41\degr13\arcmin47\farcs4 & 1.2 \\
0505720401& 00$^{\rm h}$43$^{\rm m}$19\fs98 & $+$41\degr13\arcmin47\farcs2 & 0.9 \\
0505720501& 00$^{\rm h}$43$^{\rm m}$19\fs92 & $+$41\degr13\arcmin47\farcs2 & 1.2 \\
0505720601& 00$^{\rm h}$43$^{\rm m}$19\fs92 & $+$41\degr13\arcmin47\farcs2 & 1.2 \\
\hline
\end{tabular}
\label{position}
\tablefoot{
\tablefoottext{a}{distance between the SSS and the optical position of \n12b,
(RA = 00$^{\rm h}$43$^{\rm m}$19\fs97,
Dec = $+$41\degr13\arcmin46\farcs3 J2000, BDS2009)}
}
\end{table}
In \xmm\ observations in January and February 2008 
starting 30 d after the optical outburst of \n12b, a bright source is found to be consistent with the 
position of the nova (see Table \ref{position}) that was not present in the \chandra\
HRC-I 
observations in December 2007 or earlier observations. On the basis of the position in the
observation 0505720301, we name the source XMMM31 J004319.9+411347.
Hardness ratio criteria 
\citep[see e.g.][]{PFH2005} indicate that the source has to be
classified as supersoft. Positional coincidence, time of X-ray outburst, and X-ray softness identify 
the new X-ray transient as SSS emission from nova \n12b. 
Already 21~days after outburst -- in the \xmm\ observation 0505720201 -- the source may be present just above
the detection limit (factor of more than 300 below outburst maximum). The \xmm\ ToO observation in 
July 2008 suffered from high background. Nevertheless, from the non-detection of the source we 
derive an upper limit for the source of more than a factor 200 below outburst maximum. Already 
$\sim$50~days earlier 
in \swift\ observation 00037719001, \n12b is no longer detected. However, the \swift\ upper limit is less
stringent.
Figure~\ref{nova_image_xray} shows 0.2--1~keV \xmm\ EPIC pn images of the nova field. 
Table \ref{observations} lists source count rates and absorption-corrected
luminosities or corresponding upper limits, respectively. We used the 
\swift\ XRT count rates and upper limits for the source given by BDS2009. 
We also comment on the brightness of 
\n12b. To derive luminosities and upper limits for the observations with too few photons for 
spectral fitting, we assumed a 60~eV
black-body spectrum with an absorption column of 1.0\hcm{21} as for to the best-fit model of
observation 0505720301 (see below and Sect. \ref{discuss}). For \xmm\ observations where the source
was bright, we used the results from the absorbed black-body modeling (see below).

Figure \ref{nova_lc_opt_xray} also shows the long-term X-ray light 
curve of \n12b. The source brightened between 21 and 30~days after the optical outburst to an
un-absorbed luminosity of about 8\ergs{38} inferred from the assumed black-body parameters.  
The following
data points show a decrease by about a factor of two within 30~days. Unfortunately, on day~60 after the
outburst the regular monitoring ended. In addition, we have upper limits from observations 170 and 222~days after the
outburst, which clearly indicate that the SSS had been switched off by that time. Extrapolating the intensity from the 
monitoring observations, the SSS may well have turned off as early as shortly after the end of the regular monitoring. 

\begin{figure*}
\begin{minipage}[t]{12.2cm}
\vspace{0pt}
\includegraphics[origin=br,width=12.2cm,bb= 78 207 500 322,angle=0,clip]{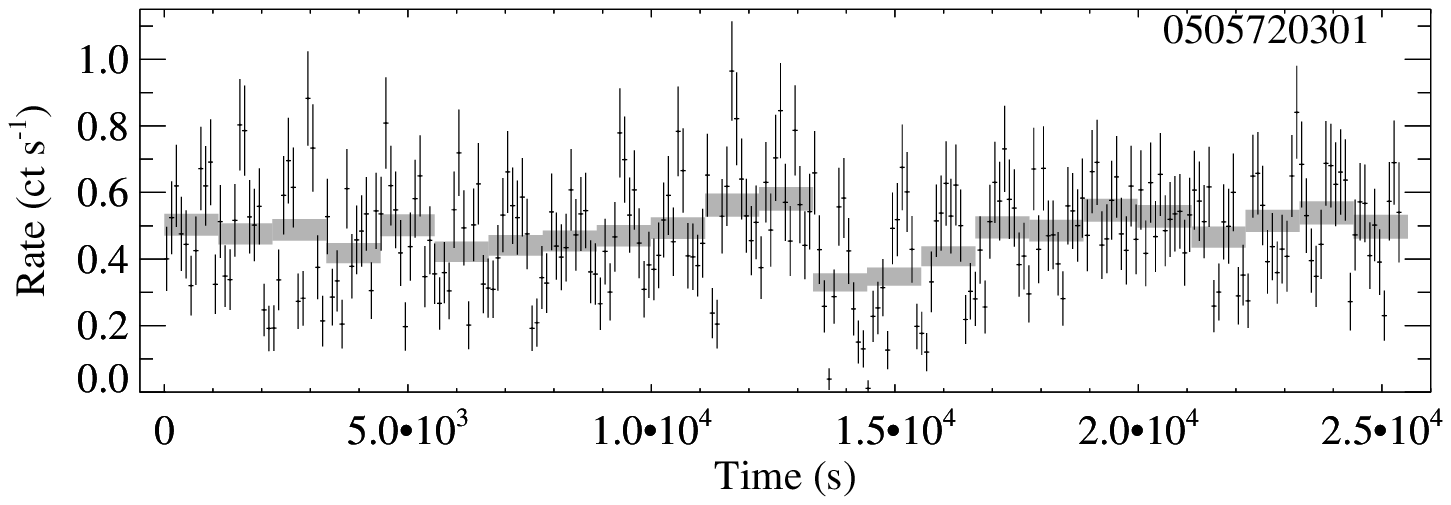} 
\end{minipage}
\begin{minipage}[t]{5.3cm}
\vspace{0pt}
\hspace{0.1cm} \includegraphics[height=5.3cm,bb= 66 35 470 699,angle=-90,clip]{16756f3b.ps}
\end{minipage}
\vfill
\begin{minipage}[t]{12.2cm}
\vspace{0pt}
\includegraphics[origin=br,width=12.2cm,bb= 78 207 500 322,angle=0,clip]{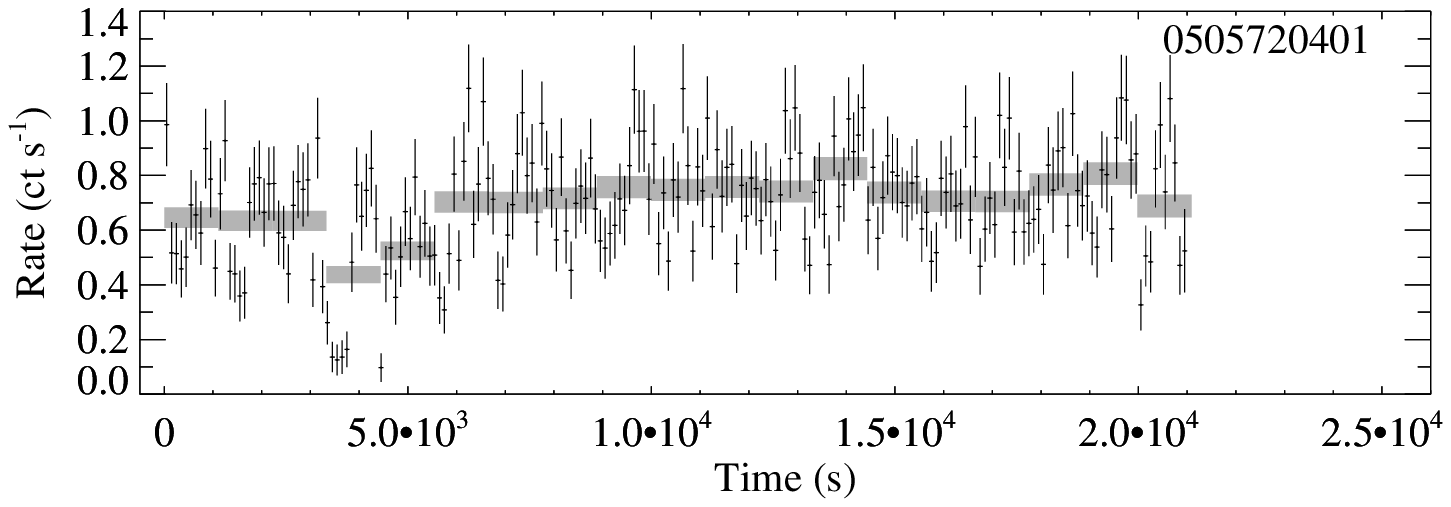} 
\end{minipage}
\begin{minipage}[t]{5.3cm}
\vspace{0pt}
\hspace{0.1cm} \includegraphics[height=5.3cm,bb=  66 35 470 699,angle=-90,clip]{16756f3d.ps}
\end{minipage}
\vfill
\begin{minipage}[t]{12.2cm}
\vspace{0pt}
\includegraphics[origin=br,width=12.2cm,bb= 78 207 500 322,angle=0,clip]{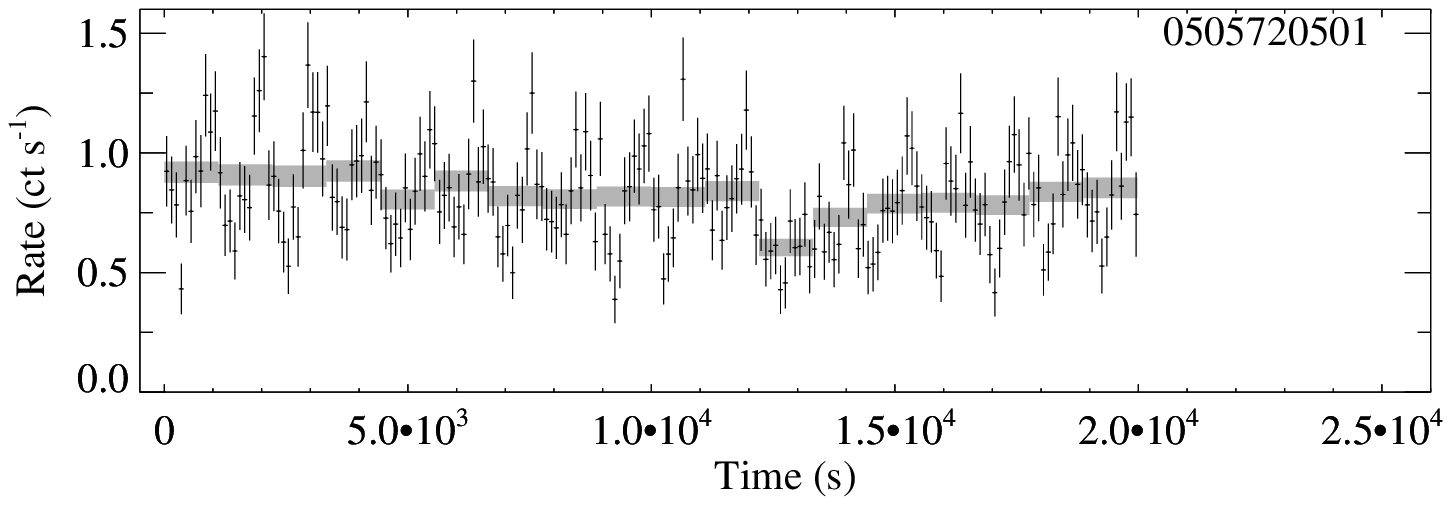} 
\end{minipage}
\begin{minipage}[t]{5.3cm}
\vspace{0pt}
\hspace{0.1cm} \includegraphics[height=5.3cm,bb=  66 35 470 699,angle=-90,clip]{16756f3f.ps}
\end{minipage}
\vfill
\begin{minipage}[t]{12.2cm}
\vspace{0pt}
\includegraphics[origin=br,width=12.2cm,bb= 78 177 500 322,angle=0,clip]{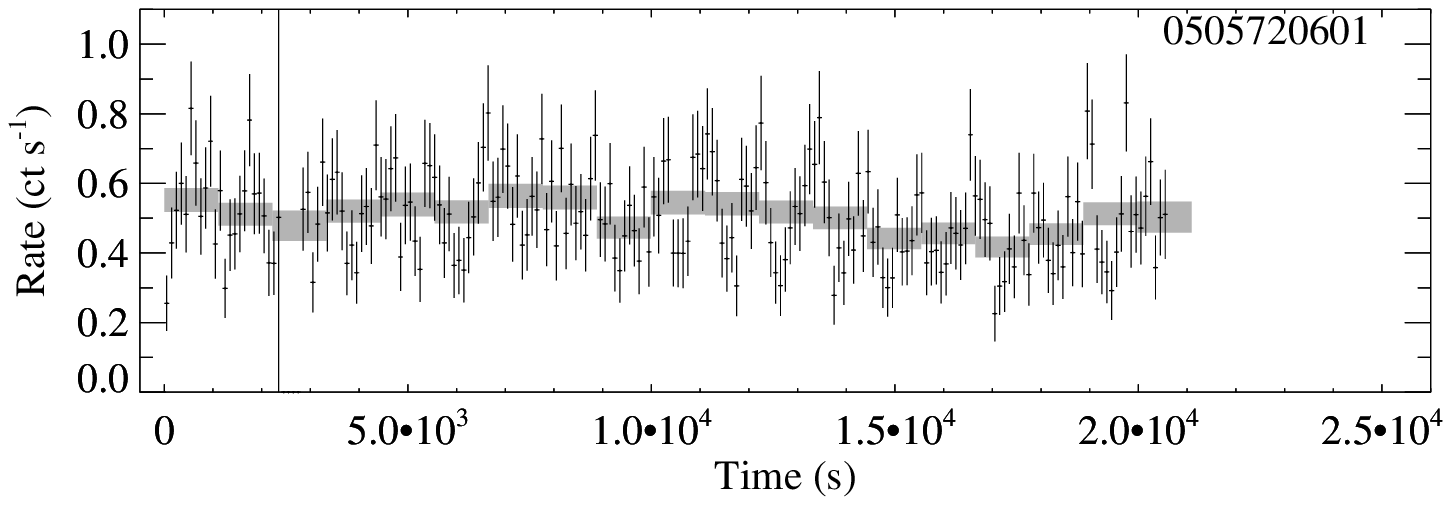} 
\end{minipage}
\begin{minipage}[t]{5.3cm}
\vspace{0pt}
\hspace{0.1cm} \includegraphics[height=5.3cm,bb=  66 35 570 699,angle=-90,clip]{16756f3h.ps}
\end{minipage}
\vfill
\caption{{\bf left:} \xmm\ EPIC pn background corrected light curves 
(0.15--1~keV) of \n12b\ for 
ObsID 0505720301, 0505720401, 0505720501, and 0505720601
integrated over 100~s (data points with error bars) and 1110~s (shaded, size indicates 1 $\sigma$ ~error).
Time zero corresponds to HJD~245\,0000.0 + (4473.2933, 4483.6332, 4492.9356,
and 4509.2042), respectively 
(solar system barycenter corrected). {\bf right:} \xmm\ EPIC pn power spectra 
(0.15--1~keV) of \n12b\ for the same observations.}
\label{fig:lc_xmm}
\end{figure*}

\begin{table}
\caption[]{Bright state \xmm\ EPIC pn timing analysis for \n12b.}
\centering
\begin{tabular}{lrrrr}
\hline\hline
\multicolumn{1}{c}{\xmm}  &
\multicolumn{1}{c}{Period} & \multicolumn{3}{c}{Pulsed fraction (band in keV)}\\ 
\multicolumn{1}{c}{ObsID}  & & 0.15--1 & 0.15--0.4 & 0.4--1 \\
& (s) & (\%) & (\%) & (\%) \\
\hline
0505720301& $1125\pm8$         & $14.7\pm3.4$ & $10.7\pm4.0$ & $21.7\pm5.6$ \\
0505720401& $1105\pm4$         & $16.7\pm2.9$ & $13.8\pm4.3$ & $18.5\pm4.1$ \\
0505720501& $1107\pm4$         & $23.0\pm2.8$ & $21.9\pm4.3$ & $24.0\pm3.7$ \\
0505720601& $1120^{+11}_{-50}$ & $14.4\pm3.5$ & $15.3\pm5.3$ & $14.1\pm5.0$ \\
\hline
\end{tabular}
\label{period}
\end{table}

\begin{figure}
   \resizebox{\hsize}{!}{\includegraphics[angle=0,clip]{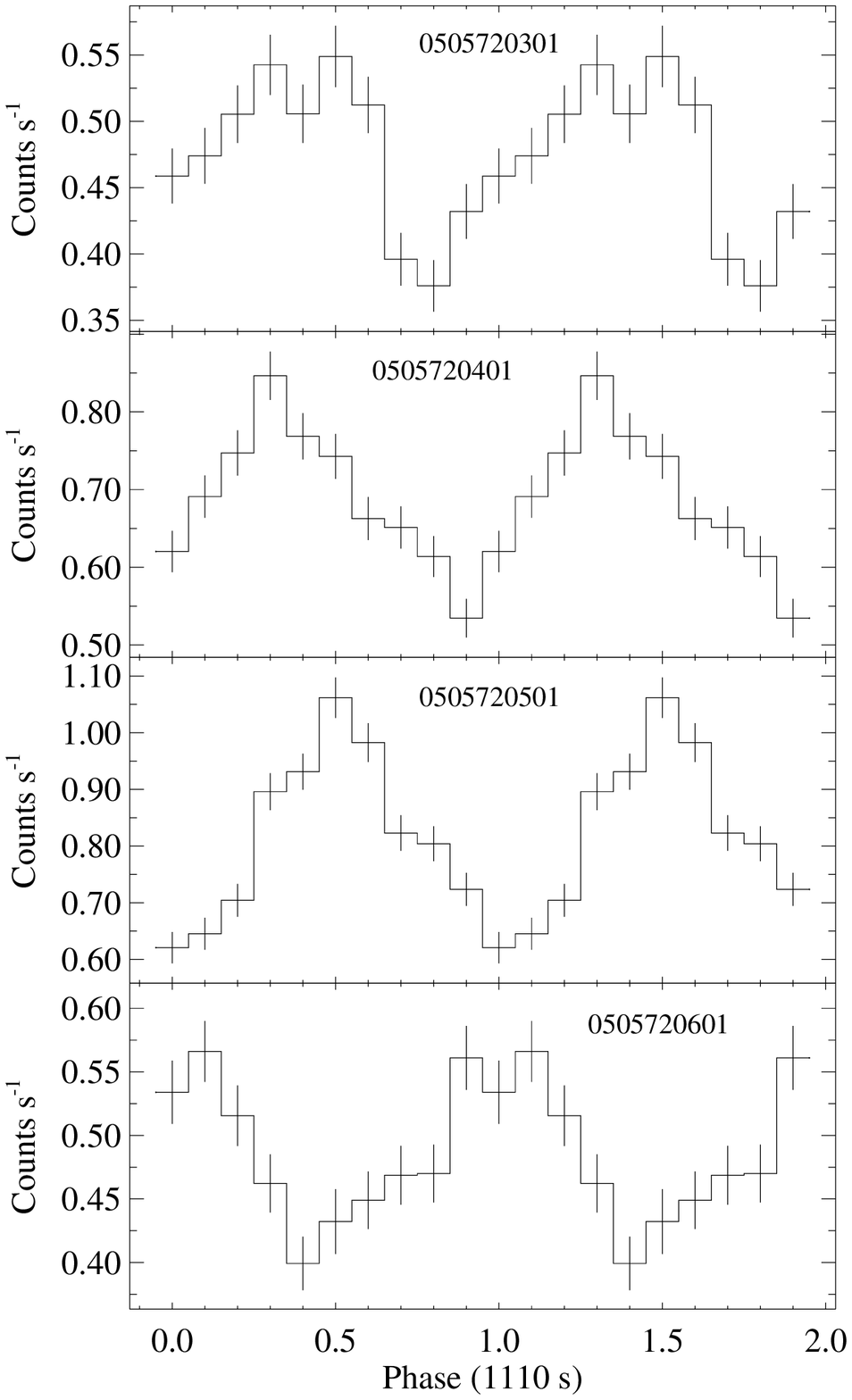}}
     \caption[]{
     Folded EPIC pn light curves in the 0.15--1~keV band. The panels show 
     the pulse profiles for the different observations.The intensity profiles 
     are background subtracted. Phase zero for all light curves corresponds 
     to  HJD~245\,4473.0.
      }
    \label{1110s_lc} 
\end{figure}

Background-corrected EPIC pn light curves of \n12b\ in the 0.15--1~keV band (Fig. \ref{fig:lc_xmm}) 
clearly show short-term time variability within the observations where \n12b\ was bright. A power spectrum
analysis revealed in all observations a significant periodicity around 1100~s. To constrain the period, we
fitted to the count rates a constant plus sinusoidal modulation. The derived periods are 
consistent within the errors with a period of 1110~s that did not vary. Figure~\ref{1110s_lc} shows the 
light curves in the 0.15--1~keV band folded modulo this period. Table~\ref{period} summarizes the 
periods found. We also give pulsed
fractions derived for the total band as well as those sub-divided into the 0.15--0.4 and 0.4--1~keV bands. 
The pulsed fraction is defined as the ratio of the amplitude of the sine function 
to the average. The quoted errors correspond to 90\% confidence intervals 
for a single interesting parameter. Pulsed fractions in the
sub-bands do not differ significantly from those of the total band. The average pulsed fraction is about 
20\%. However, there seems to be a slight variation in the pulsed fraction between observations.

To search for residual non-periodic variability in the observations, we superimpose in  Fig. \ref{fig:lc_xmm}
the data combined in 1110~s bins. Observations 0505720301, 0505720401, and 0505720501 clearly show dips 
starting at
JD 2450000.0 + (4473.4475, 4483.6717, 4493.0769), respectively. There is no significant hardness
ratio variation accompanied by these dips based on the sub-bands mentioned above. 

\begin{table*}
\caption[]{Bright state \xmm\ EPIC pn spectral best fit parameters and derived parameters for \n12b. }
\centering
\begin{tabular}{rrrrrrrrrrrr}
\hline\hline
\multicolumn{1}{c}{ObsID} & \multicolumn{1}{c}{$t_{\rm int}$} & 
\multicolumn{1}{c}{Rate\tablefootmark{a}} & \multicolumn{1}{c}{Spectra\tablefootmark{b}} &
\multicolumn{1}{c}{$N_{\rm H}$\tablefootmark{c}} & \multicolumn{1}{c}{$kT$}  & 
\multicolumn{1}{c}{$\nu$} & \multicolumn{1}{c}{$\chi^2/\nu$} &
\multicolumn{1}{c}{$L_x$\tablefootmark{d}} & \multicolumn{1}{c}{$L_{bol}$\tablefootmark{e}} & 
\multicolumn{1}{c}{$L_{bol}/L_{\sun}$} &\multicolumn{1}{c}{$R$\tablefootmark{f}}\\ 
& (ks) & ($10^{-3}$ \cts) & &(\ohcm{21}) & (eV) & & & \multicolumn{2}{c}{(\oergs{38})} & 
($10^4$) & ($10^9$ cm)\\
\hline
0505720301&  22.08 & 260$\pm$3 & BB   & $1.33_{-0.14}^{+0.15}$ & $57.7_{-2.1}^{+2.2}$ &  78 & 1.12 & 7.7 & 11.0 & 28.8 & 2.78 \\
          &        &           & WD H & $0.70_{-0.07}^{+0.08}$ & $61.0_{-0.3}^{+0.3}$ &  78 & 1.36 & 1.9 &  2.3 &  5.9 & 1.12 \\
          &        &           & WD S & $0.61_{-0.06}^{+0.06}$ & $69.4_{-0.2}^{+0.2}$ &  78 & 1.28 & 1.6 &  1.8 &  4.8 & 0.78 \\
0505720401&  17.94 & 395$\pm$5 & BB   & $1.50_{-0.14}^{+0.19}$ & $76.2_{-2.8}^{+2.4}$ &  96 & 1.63 & 6.1 &  8.4 & 21.9 & 1.31 \\
          &        &           & WD H & $0.85_{-0.06}^{+0.07}$ & $70.0_{-0.3}^{+0.3}$ &  96 & 1.47 & 2.4 &  2.7 &  7.2 & 0.94 \\
          &        &           & WD S & $0.95_{-0.11}^{+0.14}$ & $75.1_{-1.3}^{+1.6}$ &  96 & 1.29 & 2.6 &  3.0 &  7.7 & 0.85 \\
0505720501&  16.86 & 479$\pm$5 & BB   & $1.81_{-0.17}^{+0.21}$ & $81.1_{-2.4}^{+2.4}$ & 112 & 2.68 & 7.6 & 10.2 & 26.6 & 1.27 \\
          &        &           & WD H & $0.83_{-0.08}^{+0.06}$ & $75.6_{-1.3}^{+1.3}$ & 112 & 1.06 & 2.4 &  2.7 &  7.0 & 0.80 \\
          &        &           & WD S & $0.86_{-0.09}^{+0.10}$ & $80.1_{-0.7}^{+0.8}$ & 112 & 0.98 & 2.4 &  2.7 &  6.9 & 0.71 \\
0505720601&  17.42 & 291$\pm$4 & BB   & $1.36_{-0.16}^{+0.21}$ & $78.4_{-3.0}^{+2.8}$ &  71 & 1.76 & 3.7 &  5.0 & 13.1 & 1.06 \\
          &        &           & WD H & $0.77_{-0.08}^{+0.09}$ & $70.9_{-0.5}^{+0.5}$ &  71 & 0.93 & 1.6 &  1.8 &  4.7 & 0.75 \\
          &        &           & WD S & $0.78_{-0.08}^{+0.10}$ & $77.8_{-1.0}^{+0.4}$ &  71 & 0.89 & 1.6 &  1.7 &  4.6 & 0.61 \\
\hline
\end{tabular}
\label{spectra}
\tablefoot{
\tablefoottext{a}{Net count rate as given in XSPEC (0.15--1~keV)}
\tablefoottext{b}{Model spectra. BB: black-body, WD H and WD S: NLTE WD atmosphere assuming Galactic halo or solar 
         abundances, respectively (see text)}
\tablefoottext{c}{Absorption with Galactic metal abundances 
         \citep{2000ApJ...542..914W}}
\tablefoottext{d}{Un-absorbed X-ray luminosity (0.15--1~keV)}
\tablefoottext{e}{Bolometric luminosity}
\tablefoottext{f}{WD radii, for WD atmosphere models calculated as for BB}
}
\end{table*}

\begin{figure}
   \resizebox{\hsize}{!}{\includegraphics[angle=-90,clip]{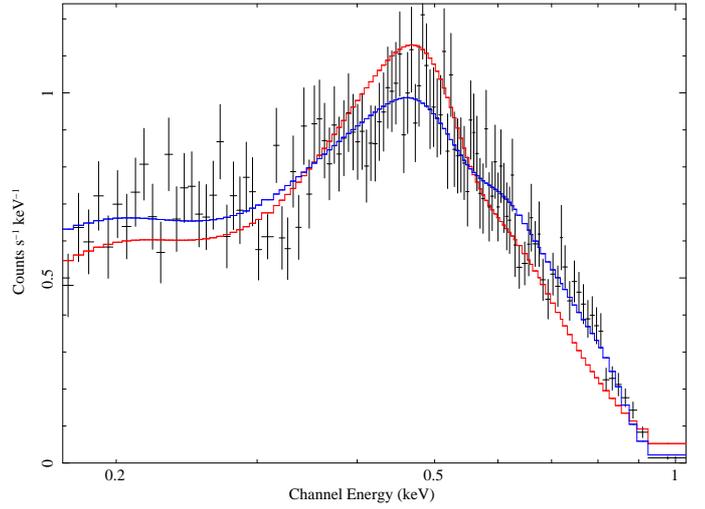}}
     \caption[]{
     \xmm\ EPIC pn spectrum of \n12b\ of ObsID 0505720501 with best-fit black-body and
     NLTE WD atmosphere (solar metal abundance) models overplotted in red and blue, respectively (see Table
     \ref{spectra}). 
     }
    \label{spec_1490} 
\end{figure}

For the four \xmm\ bright state observations of \n12b, we analyzed the X-ray spectrum following the 
procedures used by \citet{2009A&A...500..769H} for the characterization of the SSS counterpart of the
nova \hbox{M31N 2007-06b} in the \m31\ globular cluster Bol 111. To check for spectral variability
during individual observations, we investigated hardness ratios resolving the observations in 1110~s bins, 
and depending on the 1110~s pulse phase. As we found no significant hardness ratio variability, we restricted 
our analysis to spectra averaged for the individual observations.
Table \ref{spectra} summarizes the best-fit model
parameters and derived parameters for black-body and WD atmosphere models.
We give the observation identification (ObsID), the effective integration time $t_{\rm int}$, the 
raw 0.15--1~keV count rate, the spectral model used, the best-fit absorption column and
temperature. The number of energy bins reduced by the number of free parameters defines
the degree of freedom $\nu$. The goodness of fit is characterized by $\chi^2/\nu$. In addition, we
give the un-absorbed X-ray luminosity in the 0.15--1~keV band, the bolometric luminosity, and the 
radius of the emission region.
Errors represent the 90\% confidence levels.

Black-body models are a very simple approximation for the SSS emission of \n12b. They give
barely acceptable fits. 
WD atmosphere models based on more physical assumptions seem to give  acceptable
fits. We used the ``grid of synthetic ionizing spectra for very hot compact stars from NLTE model
atmospheres" computed by \citet{2003A&A...403..709R}  assuming plane-parallel geometry  and
hydrostatic and radiative equilibrium. 
The models contain all elements from H to Ca
\citep{1997A&A...320..237R}. 
\citet{2003A&A...403..709R} used the T\"ubingen Model-Atmosphere
Package \citep[TMAP,][]{2003ASPC..288..103R} with elemental abundances fixed to either
Galactic halo ([H]\footnote{[]: log(abundance/solar abundance)} = [He] = 0, [metals] = -1) or
solar ratios. The grids of model atmosphere fluxes, as well as FITS tables that can be
used in XSPEC, are available on-line\footnote{http://astro.uni-tuebingen.de/$\sim$rauch/}.
The XSPEC tables contain temperatures and fluxes for fixed surface gravity ($\log g$) and
elemental abundances. In our case, the available grid parameter space was restricted to
models with $\log g = 9.0$, since only these tables include temperatures high enough to
fit our spectra. This restriction clearly limits the significance of our best-fit solutions, since
models with different surface gravity may have provided equally good fits with different
parameters. We also note that the assumption of a plane-parallel and static atmosphere is
not physically realistic  for a nova. Furthermore, the spectral analysis is limited by the
low energy resolution of the EPIC pn spectra. Great caution should therefore be applied
when interpreting the results of the fits even when they are formally acceptable. In Fig.
\ref{spec_1490}, we show the EPIC pn spectrum of observation 0505720501 with the best-fit 
black-body and solar-abundance WD atmosphere models overplotted.

For the black-body model, bolometric luminosities and radii for the emission region
directly follow from the normalization parameter of the XSPEC fit. To derive the same
parameters for the WD atmosphere fits, we integrated the models across the range 0.005--1.0~keV, the
maximum range defined in the XSPEC tables. By comparison with the black-body models,
this may lead to an underestimation of the bolometric luminosity by about 1\%. The
corresponding radii were calculated based on the bolometric luminosity and temperature 
using formulae as for the black-body models.

The absorbing column needed for the best-fit black-body models is significantly above 
the Galactic foreground value \citep[$N_{\rm H}$ = 6.6\hcm{20},][]{1992ApJS...79...77S}
and varies between observations.
On the other hand, the best-fit absorbed WD 
atmosphere models are compatible with the Galactic
foreground column. All models indicate a temperature increase of about $\sim$15~eV
after observation 0505720301. The bolometric luminosities derived from the
WD atmosphere model fits for solar abundances indicate that the luminosity remained high
till about day~50 after outburst and then declined (see Fig.~\ref{nova_lc_opt_xray}).
The slightly lower luminosity measured at day~21 (ObsID 0505720301) may be caused by a too
small correction for foreground absorption (the best-fit $N_{\rm H}$ value is below
the Galactic absorbing column).

\section{Discussion}\label{discuss}
On the basis of the positional agreement, the short delay of the X-ray outburst after the
optical  outburst and the supersoft X-ray spectrum, the emission of XMMM31
J004319.9+411347 has to originate from the nova system \n12b. 
The speed class of the optical light curve, the decay 
of the UV emission at the onset of the SSS emission, and the start and duration of the
SSS phase are consistent with the timescales expected from the universal decline law for
classical novae \citep[see][]{2006ApJS..167...59H,2010ApJ...709..680H}. 

We have fitted absorbed black-body and WD 
atmosphere models. As discussed below, absorbed black-body models seem not to present a physically
correct description of the spectra. However, as they are often used to describe other
low resolution data we included them for comparison.  The peak luminosity of the black-body fits
significantly exceeds the Eddington luminosity of a hydrogen-rich atmosphere of a WD 
\hbox{(1.3$\times\left( \frac{M}{M_{\odot}}\right)$\ergs{38}).} In addition, the absorbing column
varies between observations and is well above the Galactic value. This confirms the 
finding that black-body fits to SSS spectra are
known to often result in too high values of $N_{\rm H}$ and too low temperatures, and
therefore point to too high luminosities  \citep[see e.g. ][ and references
therein]{1991A&A...246L..17G,1997ARA&A..35...69K}. 

The WD atmosphere models give
physically more plausible results. As indicated by our spectral fits, the resulting 
$N_{\rm H}$ values are compatible with the Galactic foreground absorption and the 
luminosity is reduced to more realistic values close to the Eddington limit. In the
following discussion, we use the results from the fits with the WD 
atmosphere model of solar abundance that corresponded to the lowest reduced $\chi^2$
values.

\subsection{\n12b system and outburst parameters}
BDS2009 determined a WD
mass in the \n12b\ system of $M_{\rm WD} \ga 1.3 M_{\sun}$ 
based on the constraints for the turn-off time for nuclear burning 
provided by one \swift\ XRT detection and one upper limit 170~d after nova outburst. 
Here we add and discuss the results from the dense \xmm\ and \chandra\ monitoring 
that allows a detailed modeling of the \n12b system and outburst parameters.
We can more tightly constrain the time of appearance and the turn-off of the SSS phase to 
$t_{\rm on} = (25\pm5)$~d and $t_{\rm off} = (115\pm55)$~d, respectively. In addition, 
we can make use of the well-determined temporal development of the X-ray temperature and luminosity.
We also discuss the radii of the X-ray emission region during the bright state derived from the EPIC 
pn spectral fits. 

To interpret our measurements, we used the envelope models for post-outburst novae from
\citet[][]{2005A&A...439.1061S}. We find an increase in the temperature to 81.5~eV in
ObsID 0505720501 and 10~d later a slight temperature decrease accompanied by a
drop in luminosity by $\sim$35\%. In view of the
model, this luminosity drop might be associated with the end of the nuclear burning. We can 
compare the \n12b\ temperature and luminosity development to 
their Table 2 and Figs. 10 and 11, which present results for five different core masses
(from 0.9 to 1.3 $M_{\sun}$) and four different chemical compositions of the WD envelope
due to mixing of the accreted material with the degenerate core. The
model parameters do not depend much for similar mixing on whether the accreting object is an ONe or a 
CO WD. 
Only models with mixing between 25\% and 50\% but closer to 25\%  of
solar-like accreted material and the degenerate core seem to explain the data. 
We note that a similar mixing was needed to model the SSS emission of the Galactic nova V1974 
Cyg \citep[][]{2005A&A...439.1057S}. While different atmosphere models change
the observed maximum temperature of $\sim$80~eV the WD by a few eV, this temperature 
determines the WD mass to
$M_{\rm WD} = (1.20\pm0.05) M_{\sun}$ (see their Fig.~7). This represents a massive WD with a
mass significantly below that derived by BDS2009. 
Our "plateau" bolometric luminosity
of about $6\times 10^4 L_{\sun}$ is slightly higher but within the errors consistent with the 
one predicted by the model ($5.2\times 10^4 L_{\sun}$). 
  
The radius of the emission region determined from the fit to the spectrum of ObsID 
0505720501 is $\sim6.4\times10^8$~cm, i.e. a factor of 1.7 above the
radius expected from a 1.2 $M_{\sun}$ non-magnetic cold WD \citep{1961ApJ...134..683H}. 
The discrepancy would be even bigger if we accept the hypothesis that the pulses in the 
X-ray light curve are caused by WD rotation.
The observed emission would have to be in-homogeneously 
distributed over the WD surface and we therefore would only measure a fraction of the 
WD surface.
However, 
during the steady hydrogen burning phase (SSS phase) one expects a larger WD photospheric 
radius, which can be derived from the photospheric temperature 
\citep[see][]{2010ApJ...709..680H}. For a temperature of 81.5~eV, we find a radius of 
$\la5.0\times10^8$~cm still slightly below our measured value. Similar radii are predicted
by the models from \citet[][]{2005A&A...439.1061S}. For a magnetic WD, even larger 
photometric radii are expected \citep[an additional radius increase by a factor of up to 1.5 
for a 1.2 $M_{\sun}$ WD,][]{2000ApJ...530..949S}. 
  
The ejected mass in the \n12b\ outburst can be estimated from both the  
time of appearance of the SSS and the expansion velocity of the ejected material. 
As in \citet[][]{2010A&A...523A..89H},
we assume the volume of the nova shell, 
expanding at constant velocity v, to be $V \sim 4\pi \times \mbox{v}^3 
\times t^3 \times f$, where the fill factor $f$ is a dimensionless 
parameter describing the thickness of the shell. We chose $f = 0.2$, 
which incorporates the thickness of the envelope due to 
thermal motions inside the gas \citep[see][and references 
therein]{2002A&A...390..155D}. Under this assumption, the column density 
of hydrogen evolves with time as $N_{H} ({\rm cm}^{-2})= M^{ej}_{H} / 
(\frac{4}{3}\pi m_H \mbox{v}^{2} t^{2} {f}')$, where 
$m_H=1.673\times10^{-24}$~g is the mass of the hydrogen atom, and $f' 
\sim 2.4$ is a geometric factor correcting the $N_{H}$ along the line of sight for
the case of a shell with fill factor $f = 0.2$. We assume that 
the expansion velocity is 2250 km s$^{-1}$ (half the FWHM  
measured in the H$\alpha$ line of the optical 
spectrum, BDS2009). We also assume that 
the SSS appears when the absorbing hydrogen column density decreases to 
$\sim10^{21}$ cm$^{-2}$. This leads to an ejected mass of
$M^{ej}_{H} = 2.0 \times 10^{-6} M_{\sun}$.

The burned mass can be estimated from the turn-off time of the SSS phase as 
$M_{\rm burn,H}=(L_{\rm bol}\cdot t_{\rm off}) / (X_{\rm H} \epsilon)$,
where $L_{\mbox{bol}}$ is the bolometric luminosity, $t_{\mbox{off}}$ the SSS turn-off time, 
$X_{\rm H}$ the hydrogen fraction of the burned material, and $\epsilon=5.98\times10^{18}$ erg g$^{-1}$ 
\citep[see][]{2005A&A...439.1061S,2010A&A...523A..89H}. From the preferred envelope mixing model of
\citet[][]{2005A&A...439.1061S} (see
above), we derive $X_{\rm H} \approx 0.5$.
Assuming the measured "plateau" 
bolometric luminosity $L_{\rm bol} = 6\times 10^4 L_{\sun}$ and that there was a drop in the X-ray luminosity after 
60~days at the end of the SSS phase, we derive $M_{\rm burn,H} = 2.0 \times 10^{-7} M_{\sun}$. However, the drop in the
X-ray luminosity after 60~days might not mark the end of the SSS phase but just represent
an intensity fluctuation during the "plateau" phase. Unfortunately, the monitoring at intervals of 10~d ended
at day~60. We therefore cannot exclude the possibility that the  \n12b\ SSS 
phase only ended shortly before day~170 after the nova outburst when the source was not detected in the \swift\ observations. 
The longer burning time would then even allow for a burned mass of $\sim 6.0 \times 10^{-7} M_{\sun}$.

\subsection{White dwarf rotation in \n12b}
The pulsations of the SSS emission of \n12b\ remained constant
within the errors at a period of 1110~s during the four observations distributed over 30~days. 
It therefore most certainly
represents the rotation period of the WD in the system. Rotation periods with similar
values have been reported for many CVs and specifically for IP systems that had a 
nova outburst \citep[see e.g.][]{2002AIPC..637....3W}. 
The 1110~s modulation is the first definite rotational period of a WD found in a \m31\ 
optical nova system. 

In CVs containing a magnetic WD, the modulation of the intensity with rotation is explained
by brighter spots on the photosphere caused by asymmetric fuel coverage due to
magnetic channeling during accretion. This may lead to steady nuclear burning close to the 
poles if enough matter is accreted producing a SSS. If the accretion is not strong enough 
-- even for a magnetic WD -- the accreted matter may not be confined to the poles until 
explosive nuclear burning sets in at the nova outburst. Therefore,
if no significant amount of new H-rich fuel arrives after the nova ignition, nuclear burning 
will quickly become spherically symmetric \citep[see e.g. discussion in][]{2002MNRAS.329L..43K}.
It is generally assumed that during the nova explosion an accretion disc in the system is 
destroyed and accretion onto the WD stops \citep[see e.g. modeling for the RN U Sco,][]
{2010ApJ...720L.195D}. However, \citet[][]{2010MNRAS.401..121P} argue that in the case of the 
nova V2491 Cyg the accretion disc was not fully destroyed and accretion resumed as early as 57~days 
after the explosion. From the lack of pulsations and other considerations, they argue that a
magnetic WD is unlikely in the system, a suggestion that had been put forward by 
\citet[][]{2009A&A...497L...5I} based on pre-nova X-ray spectra of V2491~Cyg, and used by 
\citet[][]{2009ApJ...694L.103H} to explain the secondary maximum in the optical light curve of the 
source.  During one \xmm\ observation of V2491 Cyg, \citet[][]{2011arXiv1103.4543N} find --
besides a dip -- oscillations with a period of 2232~s that are not present during the dip minimum
and also not in a second observation.
To explain the pulsations of \n12b, strong accretion onto a magnetic WD in the system would have 
had to have set in as early as 30~days after the explosion.  

Besides \n12b, only one other \m31\ nova system has been found to display a periodic 
modulation of
its X-ray emission. \citet[][]{2010A&A...523A..89H} report a $\sim$5900~s period for the SSS 
counterpart of \hbox{M31N~2006-04a}. Owing to the duration of the observation, only three cycles could be
followed. It is unclear whether the period reflects the rotation period of the WD in the system or 
is connected to the binary orbit. 

Periodicities in SSS emission from WDs have been discussed for several Galactic nova 
systems. \citet{2003ApJ...584..448D} discuss $\sim$2500~s peaks in \chandra\ observations
of the classical nova V1494~Aql as non-radial $g^+$ modes from the pulsating, rekindled WD.
Strong modulations with $\sim$35~s have been reported for the RN RS~Oph based on selected 
\swift\ XRT \citep{2006ATel..770....1O} and \chandra\ observations 
\citep{2007ApJ...665.1334N} as a transient phenomenon in the supersoft emission during 
the outburst. \citet{2008ApJ...675L..93S} report a 10700~s periodicity in \xmm\
observations of the SSS emission of nova V5116~Sgr, which they explain by partial eclipses 
by an asymmetric accretion disc. For the Galactic nova V4743~Sgr, 
\citet{2010MNRAS.405.2668D} re-analyzed \chandra\
and \xmm\ observations and identified  $\sim$1330~s as the WD rotation period.

In \m31, periodicities have been reported for two other SSS. \citet[][]{2001A&A...378..800O}
detected 865~s pulsations from the supersoft transient source XMMU~J004319.4+411759
in the center area of \m31, which was only
detected in one \xmm\ observation in June 2000. The authors speculate that the emission could
originate from the SSS phase of an optical nova for which the optical outburst was not detected.
\citet[][]{2002MNRAS.329L..43K} rejected this possibility and postulated that it represents a supersoft IP
in \m31. In view of our findings, the SSS phase of an optical nova in \m31\ still seems feasible.
\citet[][]{2008ApJ...676.1218T} detected 217~s pulsations from the bright SSS XMMU~J004252.5+411540.
The source is detected as a steady, bright SSS close to the center of \m31\  starting with the
observations of the \ein\ observatory. It most likely represents a CV
system with steady hydrogen-burning on the surface of the WD similar to the systems detected by ROSAT
in the Magellanic Clouds \citep[][]{1991Natur.349..579T,1991A&A...246L..17G}.

\subsection{\n12b: an intermediate polar (IP)?}
The dips in the \n12b\ light curves may have been caused by occulting material in the binary system and 
therefore reflect the orbital period of the system. 
Since we see a dip in three out of four observations but never more than one (see above), the
orbital period should be close to the duration of our observations. 
We can determine the possible orbital period as follows.
The time difference between the
start time of the dips in ObsIDs 0505720301 and 0505720401, as well as 0505720401 and 0505720501 are
10.2242(128)~d and 9.4052(128)~d, respectively. The differences must be multiples of the orbital
period. In addition, the difference between these differences (0.8190(256)~d) has to be a multiple n of the
period. These conditions can be fulfilled for n=2 and n=4, suggesting periods of 0.4095~d and
0.2047~d (9.828~h and 4.914~h), respectively. Shorter periods are not allowed as they would have 
lead to a second dip within ObsID 0505720401. 
Extrapolating both allowed periods to ObsID 0505720601, we would predict 
the start of a dip at HJD 2454509.3006 (corresponding to 8500~s in Fig.~\ref{fig:lc_xmm}). 
A weak dip may be present at around that time.
The proposed periods contradict the identification of a red giant secondary in \n12b\ by
BDS2009 which would point to a significantly longer orbital period of the system 
\citep[see discussion in Sect. 1 and][]{2008ASPC..401...31A}. 

While the 1110~s periodicity is deduced from many cycles and is clearly detected in four 
epochs, the orbital period derived from three dips in four separated observations, is 
less certain. We cannot exclude that this variability could also be produced by e.g. 
variations in accretion rate or occultation due to non-radial motion 
in the ejecta. However, these effects would be expected to cause more continuous variability and
would not normally create dips. However, several IP systems display dips
in their X-ray light curve that reflect their orbital period 
\citep[see e.g. light curves for the IP H2252-035/AO~Psc,][which also showed
pulsations during the dips]{1987Ap&SS.130..281P}.
Orbital periods of 4.9~h or even 9.8~h as indicated by the dips, are on the long side of -- but well
within --  the distribution of orbital periods of classical Galactic novae \citep[see
e.g.][]{1997A&A...322..807D}.  Based on its proposed rotation period of 1110~s and the 
significantly longer binary period, \n12b\ can be classified as an IP system. 
Systems with synchronized rotation and binary period (polars) may be de-synchronized 
in a nova outburst. However, rotation and binary periods should probably not differ by the amount observed 
for \n12b\ \citep[][]{2002AIPC..637....3W}.  The IP 
interpretation is also supported by the derived radius for the WD
photosphere being consistent with that of a magnetic WD. Our observations add \n12b\ as the first
extragalactic IP candidate identified with an optical nova
to the list of Galactic nova IPs and IP candidates \citep{2002AIPC..637....3W}\footnote{For an up-to-date 
catalog of IPs and IP candidates see  
http://asd.gsfc.nasa.gov/Koji.Mukai/iphome/catalog/alpha.html}. 

\section{Summary and conclusions}
During our \xmm/\chandra\ monitoring program of the central area of \m31\ to detect SSS
emission from optical novae, we have observed the outburst of the He/N nova \n12b\ in several
observations with 10~d spacing. In addition, we have analyzed a \xmm\ target of opportunity 
observation and included informations from \swift\ XRT  observations. We have performed a
source detection, determined the long-term time and spectral variations of \n12b, and searched
for shorter-term time variability in the individual observations when the source was
bright, using fast Fourier and folding techniques to analyze periodicities. 

The SSS
emission started as early as between 21 and 30~d after the optical outburst and ended between 60 and
120~d after outburst, implying that the nova \n12b\ has one of the shortest SSS phases known.
The spectrum was supersoft and can be fitted with a WD atmosphere model
with solar abundances absorbed by the Galactic foreground. The temperature of the WD
atmosphere seems to increase at the beginning of the outburst from $\sim$70 to $\sim$80~eV. 
The luminosity of \n12b\ during maximum was at the Eddington limit of a massive WD. It 
dropped by $\sim$30\% in the observation 60~d after outburst. The radius of the emission
region is $\sim6\times 10^{8}$~cm. Nova envelope models with $\la$50\% mixing between solar-like accreted material and the 
degenerate core of the WD can be used to describe the data.
We derived a WD mass of 1.2~M$_{\sun}$. From the start time of the SSS emission we
determined the ejected mass to be $2.0\times 10^{-6} M_{\sun}$. If we assume that the drop in
luminosity 60~d after outburst indicates the end of the hydrogen burning phase, we can
determine the burned mass to be $2\times 10^{-7} M_{\sun}$. The radius of the emission region 
is larger than the photospheric radius expected for a non-magnetic WD in a nova 
system but is in the range expected for magnetic WDs.

In the four bright state 
observations, we detected a stable 1110 s pulsation that we interpreted as the WD
rotation period. In addition, we detected dips in three observations that might point to a 
4.9~h or 9.8~h binary period of the system. On the basis of these periodicities together with 
the indication of a magnetic WD from the photospheric radius, we propose that \n12b\
erupted in an IP system in \m31.	

\begin{acknowledgements}
We wish to thank the referee, Matthew Darnley, for
his constructive comments, which helped to improve the clarity of the paper. 
We acknowledge the use of public data from the \swift\ data archive.
Part of this work was supported by the 
\emph{Son\-der\-for\-schungs\-be\-reich, SFB\/} 375
of the \emph{Deut\-sche For\-schungs\-ge\-mein\-schaft, DFG\/}.
The \xmm\ project is supported by the Bundesministerium f\"{u}r
Bildung und Forschung / Deutsches Zentrum f\"{u}r Luft- und Raumfahrt 
(BMBF/DLR) and the Max-Planck Society. M. Henze acknowledges support from the BMWI/DLR,
FKZ 50 OR 0405. M. Hernanz acknowledges support from grants AYA2008-01839 and 2009-SGR-315. 
G.S. acknowledges MICINN grants AYA2008-04211-C02-01 and AYA2010-15685, and 
funds from ESF EUROCORES Program EuroGENESIS through grant EUI2009-04167.
\end{acknowledgements}

\bibliographystyle{aa}
\bibliography{./paper,/home/wnp/data1/papers/my1990,/home/wnp/data1/papers/my2000,/home/wnp/data1/papers/my2001}

\end{document}